\documentclass[twocolumn,aps,pra,groupedaddress,showpacs]{revtex4}
\usepackage{epsfig,amssymb}
\def\comment#1{}\def\labell#1{\label{#1}}
\begin{document}
\title{Broadband channel capacities}
\author{Vittorio Giovannetti$^1$, Seth Lloyd$^{1,2}$, Lorenzo
Maccone$^1$, and Peter W. Shor$^3$} 
\affiliation{$^1$Massachusetts Institute of Technology --
Research Laboratory of Electronics\\ 50 Vassar st., Cambridge, MA
02139, USA\\$^2$Massachusetts Institute of Technology -- Department of
Mechanical Engineering\\ 77 Massachusetts Ave., Cambridge, MA 02139,
USA\\ $^3$\mbox{AT\&T Labs - Research 180 Park Ave, Florham Park, NJ
07932, USA}}
\date{July 14, 2003}

\begin{abstract}
  We study the communication capacities of bosonic broadband channels
  in the presence of different sources of noise. In particular we
  analyze lossy channels in presence of white noise and thermal bath.
  In this context, we provide a numerical solution for the
  entanglement assisted capacity and upper and lower bounds for the
  classical and quantum capacities.
\end{abstract}
\maketitle

The study of the broadband bosonic communication channel has been one
of the first applications of quantum communication theory
{\cite{yuen,caves}}. The basic result of this effort has been the
determination of the ultimate limits posed by quantum mechanics to the
rate at which classical information can be reliably transmitted
through the channel in the noiseless case. In this context, the
classical capacity $C$ was shown to be proportional to the square root
of the input power.  Here we generalize these results by extending the
analysis to noisy configurations and to other channel capacities such
as the quantum capacity $Q$ (the amount of quantum information that
can be reliably sent through the channel) or the entanglement assisted
capacity $C_E$ (the amount of classical information that can be
reliably sent through the channel in the presence of an infinite
amount of prior entanglement between sender and receiver).  In
particular, we study various types of noise: loss (where there is a
probability $1-\eta$ that a photon is lost in the transmission line),
loss with white noise or thermal reservoir coupling, and a
``dephasing'' channel (in which the average number of photons is
preserved, but some phase correlations are lost in the transmission).
In this context we determine the value of $C_E$ as a function of the
input power and show that the square root dependence applies also to
most of these channels. For the other capacities we provide some
bounds that establish the same dependence.  A sketch of the results
obtained is summarized in table~\ref{t:cap}: even though implicit
equations for all the capacities (or their lower bounds) have been
obtained, in most of the cases numerical methods have been employed to
derive their values as a function of the channel parameters.

\begin{table}\begin{tabular}{p{2cm}|c|c|c|}
&$C_E$ & $C$ {\scriptsize (lower bound)}  &$Q$ {\scriptsize (lower
  bound)} \\\hline 
Loss& {\em numeric}&\em analytic & \em numeric\\\hline
White noise& \em numeric&\em analytic & \em numeric
\\\hline
Thermal
reservoir&\em numeric&\em analytic &
\em numeric\\
\hline Dephasing channel& \em numeric& \em numeric &
\em numeric\\\hline \end{tabular}\caption{Capacities
calculated in the paper for different noise models.\labell{t:cap}}
\end{table}

We start by introducing the model of the channel and of its noise
sources in Sec.~\ref{s:bosonic}. We introduce capacities $C_E$, $C$
and $Q$ and the Lagrange procedure that is used to evaluate them for
the broadband channel in Sec.~\ref{s:cap}.  The remaining sections are
devoted to the analysis of the lossy channel (Sec.~\ref{s:lossce}),
the white noise channel (Sec.~\ref{s:white}), the thermal noise
channel (Sec.~\ref{s:thce}) and the dephasing channel
(Sec.~\ref{s:depch}).

\section{Gaussian bosonic channel}\labell{s:bosonic}
The prototype of a high capacity communication channel is an optical
fiber, where time or frequency multiplexing (or hybrid strategies) are
used to send information. From a fundamental point of view, such a
communication line is described as a broadband bosonic channel
{\cite{caves}}.  In the present paper we analyze the performance of
this channel in the realistic scenario of non-perfect transmissivity,
i.e. the possibility that photons can be lost during the communication
or that they can be replaced by photons coming from external noise
sources. The analysis is complicated by the fact that, for some
capacities, it is not known whether the additivity property holds,
i.e. whether entangling successive uses of a noisy channel may
increase its transmission rate {\cite{shor}}.

Without loss of generality, we will assume that for each frequency in
the channel only one polarization is used to transmit information,
i.e. no frequency degeneracy is present. The quantum description of
this channel is obtained by coupling each mode to a noise reservoir
with beam splitters that have transmissivity equal to the quantum
efficiency $\eta_j$ of the $j$th mode, i.e.
\begin{eqnarray}
a'_j=\sqrt{\eta_j}\;a_j +\sqrt{1-\eta_j}\;b_j
\;\labell{ch},
\end{eqnarray}
where $a_j$, $a'_j$ and $b_j$ are the annihilation operators of the
input, output and noise modes respectively.  The loss map ${\cal N}_j$
for the $j$th mode arises by tracing away the noise mode $b_j$ and the
global loss map $\cal N$ is the tensor product   $\bigotimes_i{\cal
N}_i$. Notice that for $\eta_j=1$ the noise reservoir is decoupled
from the transmission line: this describes a noiseless channel where
${\cal N}=\openone$. Different types of noise can be described
depending on the initial state of the reservoir modes $b_j$. We will
analyze the case in which the reservoir is in a separable Gaussian
state of the form $\rho^{(b)}=\otimes_i\;\rho_i^{(b)}$ with
 \begin{eqnarray}
  \rho^{(b)}_j=\frac\hbar{2\pi}\int\!\! dz\;\exp
\left[-i\;z\cdot\left(\matrix{\Delta q_j\cr\Delta
    p_j}\right)- \frac{z\cdot B_j\cdot z^T}2\right]
\;\labell{bathgaus},
\end{eqnarray}
where $z$ is a real bidimensional line vector, $\Delta q_j=q_j-\langle
q_j\rangle$, $\Delta p_j=p_j-\langle p_j\rangle$, with $q_j$ and $p_j$
the quadratures $q_j=\sqrt{\hbar/2}\;(b_j^\dag+b_j)$ and
$p_j=i\sqrt{\hbar/2}\;(b_j^\dag-b_j)$.  For the situations in which we
are interested, $\langle q_j\rangle=\langle p_j\rangle=0$ and the
correlation matrix $B_j$ in  Eq.~(\ref{bathgaus}) has the form
\begin{eqnarray}
B_j=\hbar
\left[\matrix{\bar{N}_j+1/2&0\cr0&\bar{N}_j+1/2}\right]
\;\labell{matriceb},
\end{eqnarray}
where $\bar N_j$ is the average number of photons in the noise mode
$b_j$.  With this choice of $\rho^{(b)}$, the CP-map $\cal N$
describes a Gaussian channel, i.e. it transforms Gaussian input states
(in the modes $a_j$) into Gaussian output states (in $a'_j$).  

Four different noise models will be analyzed in detail in this paper.
The simplest one is a purely lossy channel in which the photons in the
$j$th mode have a probability $1-\eta_j$ to be lost during
transmission. It is described by taking $\bar N_j=0$ for all $j$, i.e.
by taking $\rho^{(b)}$ in the vacuum.  For optical communications this
is the most interesting situation, since thermal photons are
negligible at room temperatures.  Another interesting case is given by
choosing $\bar N_j=\bar N$ for all $j$, which describes an added white
noise to the transmission. On the other hand, by choosing
  \begin{eqnarray} \bar N_j=\frac 1{e^{\hbar\omega_j/{(KT)}}-1}
\;\labell{bose},
\end{eqnarray} (with $\omega_j$ the $j$th mode frequency) we can
describe the effect of coupling the communication line to a thermal
reservoir at temperature $T$. Of course, in the limits $\bar N\to 0$
or $T\to 0$ both the white noise and the thermal channel reproduce the
lossy channel. The common trait among these three noise models is the
fact that they can be parametrized as
  \begin{eqnarray} \bar N_j=\bar N\;v(\omega_j/\bar\omega)
\;\labell{caratt},
\end{eqnarray}
where $\bar N$ describes a characteristic number of photons in the
transmission and $\bar\omega$ describes (through an appropriate
function $v$) an eventual characteristic frequency of the channel. The
parametrization (\ref{caratt}) will be useful in deriving some scaling
properties that simplify the derivation. A final noise model we will
analyze is a non-linear noise mechanism where the average photon
number of the reservoir $\bar N_j$ is a function of the average photon
number in the message. This type of model is well suited to describe
situations in which the noise is due to the action of some active
third party (e.g. an eavesdropper) who is tampering with the
transmission. In particular we will analyze a sort of dephasing
channel where the average photon number in each transmission mode is
preserved, even though some phase correlation is lost.  Because of the
non-linearity of this noise, the parametrization (\ref{caratt}) does
not apply to the dephasing channel, but most of the general formalism
developed for the other models can still be used in this case.

\section{Capacities}\labell{s:cap}
In this section we introduce the three different channel capacities
that will be analyzed in the paper.
\paragraph*{Entanglement assisted capacity.---}\labell{s:ce}
The one channel capacity that is known to be additive
{\cite{shorprl,shor2}} even in the presence of noise is the
entanglement assisted capacity $C_E$. It is defined as the number of
bits that can be reliably transmitted per channel use in the presence
of an unlimited quantity of prior entanglement shared among the sender
and the receiver. This quantity gives a simple upper bound to all the
other channel capacities and is conjectured to provide an equivalence
class for channels {\cite{shor2}}. Analogously, one can define the
entanglement assisted quantum capacity $Q_E$ that measures the number
of {\it qubits} that can be reliably transmitted per channel use in
the presence of an unlimited quantity of prior entanglement. Using
teleportation and superdense coding, it is easy to show that
$Q_E=C_E/2$, so that only one of these two quantities needs to be
determined {\cite{shorprl}}.

Taking advantage of its additivity property, the entanglement assisted
capacity of a multimode channel can be calculated as
{\cite{shor2,holevoce,paper2}}
\begin{eqnarray} 
C_E=\max_{\varrho_{j}\in{\cal H}_{j}}
\Big\{\sum_{i}I({\cal N}_i,\varrho_i)\Big\} 
\;\labell{addit},
\end{eqnarray}
where ${\cal H}_j$ is the Hilbert space of the $j$th mode in the
channel and $I({\cal N}_j,\varrho_j)$ is the quantum mutual
information defined as  {\cite{cerfadami}}
\begin{eqnarray}
I({\cal N}_j,\varrho_j)\equiv S(\varrho_j)+S({\cal N}_j[\varrho_j])-S(({\cal
N}_j\otimes\openone)[\Phi_{\varrho_j}])
\;\labell{qmi},
\end{eqnarray}
(with $S(\varrho_j)=-$Tr$[\varrho_j\log_2\varrho_j]$ the Von Neumann 
entropy and $\Phi_{\varrho_j}$ a purification of the mode input
density matrix $\varrho_j$). The
maximization (\ref{addit}) will be performed only using the states
$\varrho_j$ that satisfy the average energy constraint
\begin{eqnarray}
\sum_{i}\hbar\omega_iN_i={\cal E}
\;\labell{energ},
\end{eqnarray}
where $\omega_j$ is the frequency of the $j$th mode and
$N_j=\mbox{Tr}[a^\dag_ja_j\varrho_j]$ is its average number of
photons.  This constraint is fundamental: without any restriction the
bosonic channel would have infinite capacity since the Hilbert space
that the sender could use for encoding would be infinite-dimensional.
The energy constraint introduces an effective cut-off in the dimension
of the coding space {\cite{caves}}. This, of course, mirrors any
realistic situation in which the energy available for the transmission
is always finite.

Since we are dealing with a Gaussian channel $\cal N$ we can apply the
Holevo-Werner theorem {\cite{holevo}} which states that $I({\cal
  N}_j,\varrho_j)$ reaches its maximum over Gaussian inputs.
Moreover, for the noise models we analyze, squeezing the input to the
$j$th mode does not increase its quantum mutual information if the
energy of the mode is fixed (see App.~\ref{s:entr} for details).
Hence, the maximum value of $I({\cal N}_j,\varrho_j)$ is given by an
expression $c_E(N_j,\bar N_j,\eta_j)$ that depends only on the number
of photons $N_j$, on the noise parameter $\bar N_j$, and on the
quantum efficiency $\eta_j$. The explicit form of $c_E$ is evaluated
in App.~\ref{s:entr} and is
\begin{eqnarray}
&& c_E(N_j,\bar{N}_j,\eta_j)=
g(N_j)+g(N'_j)\;\labell{cdiewhite}\\\nonumber&&
-g\left(\frac{D_j+N_j-N_j'-1}2\right)-
g\left(\frac{D_j-N_j+N_j'-1}2\right)\;,
\end{eqnarray}
where
\begin{eqnarray}
N_j'&\equiv&\eta_jN_j+(1-\eta_j)\bar N_j\;\labell{nprimo}.
\end{eqnarray}
is the average photon number in the $j$th mode at the channel output and
\begin{eqnarray}
D_j&\equiv&\sqrt{(N_j+N'_j+1)^2-4\eta_jN_j(N_j+1)}\;\labell{dj},\\
g(x)&\equiv&\left\{\begin{array}{ll}(x+1)\log_2(x+1)-x\log_2(x)& 
      \mbox{ for }x\neq
      0\cr 0&\mbox{ for } x=0\;.\end{array}\right.\nonumber\;\\\labell{defdig}
\end{eqnarray}
In terms of $c_E$, Eq.~(\ref{addit}) becomes   \begin{eqnarray}
C_E=\max_{N_{j}} \Big\{\sum_{i}c_E({N}_i,\bar N_i,\eta_i)\Big\}
\;\labell{addit2},
\end{eqnarray}
where the maximization must be performed on the $N_j$'s that satisfy
(\ref{energ}). In Sec.~\ref{s:lag} we will calculate explicitly the
right hand side of (\ref{addit2}). First, however, it is convenient to
introduce the other channel capacities in order to underline some
common features.

\paragraph*{Classical capacity.---}\labell{s:ccq}
The classical capacity $C$ measures the quantity of bits that can be
sent reliably through the channel per channel use (without assistance
of prior entanglement). For the noiseless broadband bosonic channel
($\eta_j=1$) it has been shown \cite{yuen,caves} that, under the
energy constraint (\ref{energ}), \begin{eqnarray}
  C=\max_{\varrho\in{\cal H}}\left\{S[\varrho]\right\}={\cal T}\;R_C
  \;\labell{classicalc},
\end{eqnarray}
where $\cal T$ is the transmission time, ${\cal H}=\otimes_i{\cal
  H}_i$ is the Hilbert space of the multimode channel (${\cal H}_j$
being the space of the $j$th mode), and
\begin{eqnarray} R_C=\frac 1{\ln 2}\sqrt{\frac{\pi{\cal
P}}{3\;\hbar}}\;\labell{erreci}
\end{eqnarray}
is the classical communication rate in terms of the input power ${\cal
  P}={\cal E}/{\cal T}$.

In the presence of noise, however, a recipe to calculate $C$
involving only single uses of the channel, as in the case of
Eqs.~(\ref{addit}) and (\ref{classicalc}), is not known. It could be
that entangling successive uses of the channel the amount of
information transmitted is increased {\cite{schumi1,holevo3}}. This
would require to consider input states in the Hilbert space ${\cal
  H}^{\otimes n}$ pertaining to $n$ successive uses of the channel.
The estimation of $C$ is, hence, a daunting task. Nevertheless, a
simple lower bound for it can be obtained considering unentangled
coding procedures, where the sender is not allowed to employ codewords
which entangle different channel uses. In the multimode channel, a
further simplification consists in considering coding procedures where
entanglement among the different signal modes $a_{j}$ is forbidden.
This yields the inequality
\begin{eqnarray}
C\geqslant\max_{p_j{(\mu)},\sigma_j{(\mu)}}\Big\{\sum_{i}
{\cal
    X}_{{\cal N}_i}\left(p_i{(\mu)},\sigma_i{(\mu)}\right)\Big\}
  \;\labell{add},
\end{eqnarray} 
where $\varrho_j=\int d\mu\;p_j{(\mu)}\sigma_j{(\mu)}$ describes a
message in which the $\mu$th ``letter'' encoded in the density
operator $\sigma_j{(\mu)}\in{\cal H}_j$ has probability density
$p_j{(\mu)}$ to be sent in the $j$th mode and where
\begin{eqnarray} {\cal X}_{{\cal
    N}_j}\equiv S({\cal
    N}_j[\varrho_j])-\int d\mu\;p_j{(\mu)}S({\cal
    N}_j[\sigma_j{(\mu)}])\;\labell{hole}
\end{eqnarray} 
is the Holevo information. Unlike the case of $C_E$ discussed in the
previous section, it is not known whether the maximum of
Eq.~(\ref{add}) can be evaluated working only with Gaussian states.
However, adopting this strategy one still obtains a tight lower bound
for $C$ {\cite{holevo}}. Thus, we evaluate ${\cal X}_{{\cal
    N}_j}(p_j{(\mu)},\sigma_j{(\mu)})$ for the $j$th mode using
coherent states $\sigma_j{(\mu)}=|\mu\rangle_j\langle\mu|$ weighted
with Gaussian probability distribution
$p_j{(\mu)}=\exp[-|\mu|^2/N_j]/(\pi N_j)$. Selecting this encoding we
are assuming that, as in the case of $C_E$, squeezing does not
increase the unassisted capacity if there is an average energy
constraint on the input state {\cite{hirota,hirota1,yuen,caves}}. With
this choice, Eq.~(\ref{add}) can be written in a form analogous to
(\ref{addit2}), i.e.
\begin{eqnarray}
C\geqslant\max_{N_j}\Big\{\sum_{i}k(N_i,\bar N_i,\eta_i)\Big\}
\;\labell{do}.
\end{eqnarray}
where the maximum must again be taken under the average energy
constraint (\ref{energ}). The function $k$ is calculated in
App.~\ref{s:entr} and is given by
\begin{eqnarray}
k(N_j,\bar N_j,\eta_i)=g(N'_j)-g((1-\eta_j)\bar N_j)
\;\labell{dk},
\end{eqnarray}
with $N'_j$ defined in Eq.~(\ref{nprimo}).  

Equation (\ref{do}) establishes a lower bound for the classical
capacity $C$.  Simple upper bounds for $C$ are given by the
entanglement assisted capacity $C_E$ of Eq.~(\ref{cidiefin}) and by
the noiseless classical capacity ${\cal T}\;R_C$ of
Eq.~(\ref{classicalc}).

\paragraph*{Quantum capacity.---}\labell{s:q}
The quantum capacity $Q$ of a channel is the number of qubits that can
be sent reliably through the channel per channel use. For the
noiseless case one can show that $Q=C$, i.e. for bosonic channel with
$\eta_j=1$ one can show that $Q={\cal T}\;R_C$. As for the classical
capacity, an expression involving only single uses of the channel is
not known for noisy channels: again it could be that the entanglement
of successive channel uses might increase $Q$ {\cite{sethq,barnum}}.
Also here we will consider the lower bound obtained by excluding all
the coding procedures that make use of entanglement among successive
uses of the channel or among different modes. This provides the
inequality
\begin{eqnarray}
Q\geqslant\max_{\varrho_j}\Big\{\sum_{i}J({\cal N}_i,\varrho_i)
\Big\}\;\labell{q},
\end{eqnarray}
where \begin{eqnarray} J({\cal N},\varrho)\equiv I({\cal
    N},\varrho)-S(\varrho)\;\labell{coher}
\end{eqnarray}
is
the coherent information {\cite{schumi,sethq}}. Equation (\ref{q}) is
a consequence of the fact that random quantum codes can convey a
number of qubits equal to the coherent information of the channel (if
it is greater than zero) {\cite{sethq}}--- for the particular case of
the Gaussian channels this same result was proved also in
{\cite{preskill}}.  In evaluating the right side of Eq.~(\ref{q}), we
will employ Gaussian states: here the Holevo-Werner theorem does not
apply and this choice will further lower the bound (\ref{q}) on $Q$.
Moreover, we will restrict the analysis to non-squeezed inputs. These
considerations allow us to write Eq.~(\ref{q}) as (see
App.~\ref{s:entr})
\begin{eqnarray} Q\geqslant\max_{{N_j}}\sum_iq(N_i,\bar N_i,\eta_i)
  \;, \;\labell{qgen}
\end{eqnarray}
where the maximization must be performed under the energy constraint
(\ref{energ}) and (see App.~\ref{s:entr}) \begin{eqnarray} q(N_j,\bar
  N_j,\eta_j)&=&g(N'_j)-g\left(\frac{D_j+N_j-N_j'-1}2\right)
\nonumber\\&&- g\left(\frac{D_j-N_j+N_j'-1}2\right)\;,
  \;  \;\labell{dq}
\end{eqnarray}
with $N'_j$ and $D_j$ defined in (\ref{nprimo}) and (\ref{dj})
respectively.

An alternative lower bound for the quantum capacity $Q$ can be
obtained by observing that the definitions of $C_E$ in (\ref{addit})
and of $I$ in (\ref{qmi}) imply
    \begin{eqnarray} C_E&=&\max_{\varrho\in{\cal H}}\Big\{J({\cal
N},\varrho)+S(\varrho)\Big\}\leqslant 
\max_{\varrho\in{\cal H}}\Big\{J({\cal
N},\varrho)\Big\}\nonumber\\&&+\max_{\varrho\in{\cal H}}
\Big\{S(\varrho)\Big\}\leqslant
Q+\max_{\varrho\in{\cal H}}\Big\{S(\varrho)\Big\} \;\labell{maggioraz},
\end{eqnarray} 
which for the broadband channel gives $Q\geqslant C_E-{\cal T}\;R_C$
by employing Eq.~(\ref{classicalc}).

Equations (\ref{qgen}) and (\ref{maggioraz}) give two lower bounds to
$Q$. A simple upper bound is given by the entanglement assisted
quantum capacity $Q_E=C_E/2$.
\begin{table*}\begin{tabular}{|c|c|}
\hline&\\$C_E$&$\displaystyle\left(1+\frac 1{N_j}\right)\left(1+\frac
1{N_j'}\right)^{\eta_j} =e^{\omega_j/\Omega}\left(1+\frac
2{D_j+N_j-N'_j-1}\right)^{{(A_j+1-\eta_j)}/2}\left(1+\frac
2{D_j-N_j+N'_j-1}\right)^{{(A_j-1+\eta_j)}/2}$ \\ & \\ \hline 
&\\ $C$  & $\displaystyle\left(1+\frac
1{\eta_jN_j+(1-\eta_j)\bar N_j}\right)^{\eta_j}=e^{\omega_j/\Omega}$
\\{\scriptsize lower bound}& \\ \hline &\\
$Q$ & $\displaystyle\left(1+\frac 1{N_j'}\right)^{\eta_j}
    =e^{\omega_j/\Omega}\left(1+\frac
      2{D_j+N_j-N'_j-1}\right)^{{(A_j+1-\eta_j)}/2}\left(1+\frac
      2{D_j-N_j+N'_j-1}\right)^{{(A_j-1+\eta_j)}/2}$ \\{\scriptsize lower
  bound}&\\ \hline  \end{tabular}\caption{Lagrange equations deriving
from (\ref{lagr}) for the different capacities. The functions $N'_j$
and $D_j$ are defined in (\ref{nprimo}) and (\ref{dj})
respectively and
$A_j\equiv[{(1-3\eta_j)N_j+(1-\eta_j)+(1+\eta_j)N'_j}]/{D_j}$. Notice
that the equation pertaining to $C$ can always be solved analytically.
\labell{t:lagr}} 
\end{table*}

\subsection{Lagrange multiplier procedure}\labell{s:lag}
In order to determine the values of $C_E$ and the lower bounds for $C$
and $Q$ given by Eqs.~(\ref{addit2}), (\ref{do}) and (\ref{qgen}), one
needs to perform maximizations of the form \begin{eqnarray}
  W=\max_{N_j}\Big\{\sum_jw(N_i,\bar N_i,\eta_i)\Big\} \;\labell{w},
\end{eqnarray}
under the constraint given by Eq.~(\ref{energ}). In Eq.~(\ref{w}), the
quantity $W$ represents $C_E$ or the lower bounds for $C$ or $Q$
depending on whether $w$ is equal to $c_E$, $k$ or $q$ respectively.
The Lagrange multiplier procedure is well suited to perform these
constrained maximizations. It amounts to finding the values of
$\{N_j\}$ which solve the equations
    \begin{eqnarray} \frac{\partial\ }{\partial N_j}\left[\sum_{i}
w(N_i,\bar N_i,\eta_i)-\frac{1}{\Omega\ln 2}\sum_{i}\omega_iN_i\right]=0
\;\labell{lagr},
\end{eqnarray}
where $1/(\Omega\ln 2)$ is the Lagrange multiplier that must be chosen
to satisfy the constraint (\ref{energ}) after having solved
Eq.~(\ref{lagr}). The explicit expressions of Eq.~(\ref{lagr}) for the
three capacities are reported in table~\ref{t:lagr}. These equations
are in general difficult to solve.  A first useful simplification is
to assume that all the modes have the same quantum efficiency, i.e.
$\eta_j=\eta$ for all $j$. Even though this is a strong assumption, it
is still a good description for broadband channels that have a wide
spectral transmission window. Under this approximation, it is easy to
verify that Eq.~(\ref{lagr}) has solution that depends on the mode
frequency $\omega_j$ and on the noise parameters $\bar\omega$ and
$\bar N$ of Eq.~(\ref{caratt}) as
\begin{eqnarray}
N_j={\cal F}\left(\frac{\omega_j}\Omega,\frac{\bar\omega}
\Omega,\bar N,\eta\right)\;\labell{ndij}.
\end{eqnarray}
To calculate $\Omega$, the energy constraint (\ref{energ}) can be
written as
\begin{eqnarray}
\frac{\cal E}\hbar&=&\sum_{i}\omega_i\;{\cal F}\left(\frac{\omega_i}\Omega,\frac{\bar\omega}
\Omega,\bar N,\eta\right)\nonumber\\
&\simeq&
\int_0^\infty \frac{d\omega}{\delta\omega}\;\omega\;{\cal
F}\left(\frac{\omega}\Omega,\frac{\bar\omega}
\Omega,\bar
N,\eta\right)\nonumber\\&=&\frac{\Omega^2}{\delta\omega}\int_0^\infty
dx\;x\;{\cal
F}\left(x,\frac{\bar\omega}
\Omega,\bar
N,\eta\right)
\;\labell{condiz},
\end{eqnarray}
where the sum over the mode index $i$ is approximated with an integral
over the mode frequencies, under the assumption of small minimum
frequency interval $\delta\omega$ of the channel. This last quantity
determines the minimum time ${\cal T}=2\pi/\delta\omega$ needed to
transmit a signal in the channel. In order to solve (\ref{condiz}) in
terms of $\Omega$, it is useful to introduce the adimensional
parameter $y_0=\bar\omega/\Omega$. Thus we find
\begin{eqnarray}
\Omega=\left[{\frac{2\pi\; {\cal P}}{\hbar\;f(y_0,\bar N,\eta)}}\right]^{1/2}
\;\labell{omegagr},
\end{eqnarray}
where ${\cal P}={\cal E}/{\cal T}$ is the average input power,
\begin{eqnarray} f(y_0,\bar N,\eta)\equiv\int_0^\infty dx\;x\;{\cal
    F}(x,y_0,\bar N,\eta)
\;\labell{defdif},
\end{eqnarray}
and $y_0$ is determined by solving (with respect to $y$) the equation
  \begin{eqnarray} y^2=\frac{\hbar\bar\omega^2}{2\pi{\cal P}}\;f(y,\bar
N,\eta) \;\labell{y}.
\end{eqnarray}
If the noise reservoir does not have a characteristic frequency
$\bar\omega$ (as in the case of the loss, white noise and dephasing),
the derivation simplifies since neither $\cal F$ nor $f$ depend on the
parameter $y_0$: Eq.~(\ref{y}) does not apply and $\Omega$ is already
determined by Eq.~(\ref{omegagr}). However, for the sake of
generality, we can include also these last cases in the above
formalism by assigning to them $\bar\omega=0$ and $y_0=0$.

The value of $W$ is finally obtained using Eqs.~(\ref{ndij}) and
(\ref{omegagr}) to evaluate the sum (\ref{w}), i.e.
    \begin{eqnarray} W&=&\sum_{i}w\left({\cal
F}\left(\frac{\omega_i}\Omega,\frac{\bar\omega}\Omega,\bar
N,\eta\right),\bar N_i,\eta\right) 
\labell{cidie1}\\\nonumber& \simeq&\int_0^\infty
\frac{d\omega}{\delta\omega}\;w\left({\cal
F}\left(\frac{\omega}\Omega,\frac{\bar\omega}\Omega,\bar
N,\eta\right),\bar N\;v\left(\frac\omega{\bar\omega}\right),\eta\right) 
\\\nonumber&=&\frac\Omega{\delta\omega}\int_0^\infty
dx\;w\left({\cal
F}\left(x,y_0,\bar
N,\eta\right),\bar N\;v\left(\frac x{y_0}\right),\eta\right) 
\;,
\end{eqnarray}
where the parametrization (\ref{caratt}) was employed.  Apart from
corrections of order $1/{\cal T}$ (coming from the approximation of
the mode sum with the frequency integral), Eqs.~(\ref{omegagr}) and
(\ref{cidie1}) imply that
\begin{eqnarray} W={\cal T}\;R_C\;{\cal W}(y_0,\bar N,\eta)
\;\labell{cidiefin},
\end{eqnarray}
where $R_C$ is the noiseless classical rate of Eq.~(\ref{erreci}) and
\begin{eqnarray} &&{\cal W}(y_0,\bar N,\eta)\equiv\frac {\ln
    2}{\pi}\sqrt{\frac 3{2\;f(y_0,\bar
      N,\eta)}}\labell{calci}\\\nonumber&& \times\int_0^\infty\!\!
  dx\;w\left({\cal F}\left(x,y_0,\bar N,\eta\right),\bar
    N\;v\left(\frac x{y_0}\right),\eta\right)\;.
\end{eqnarray}
The quantity $\cal W$ is a proportionality factor that characterizes
the dependence of $W$ on the noise parameters $\eta$, $\bar N$ and
$\bar\omega$.  Even though it is in general difficult, if not
impossible, to analytically evaluate the expressions $\cal F$ and
${\cal W}$, one can still provide numerical solutions for these two
quantities as will be shown in the following subsections.

When there is no characteristic frequency $\bar\omega$ (as in the
cases of loss, white noise and dephasing), then $y_0=0$ and
Eq.~(\ref{cidiefin}) tells us that $W$ depends on the input power
$\cal P$ only through the classical communication rate $R_C$. This
means that, in these cases, the capacities of the channel (or at least
their bounds) are proportional to the square root of $\cal P$ just as
the noiseless channel classical capacity of Eq.~(\ref{classicalc}). On
the other hand, when a characteristic frequency $\bar\omega$ does
exist (as in the case of the thermal noise), then $W$ depends on $\cal
P$ also through the parameter $y_0$, which, according to Eq.~(\ref{y})
is a non-trivial function of $\bar\omega^2/{\cal P}$. However, for
fixed value of this ratio, the $\sqrt{\cal P}$ proportionality still
applies.

\section{Lossy channel}\labell{s:lossce}
The simplest channel is the lossy channel where the reservoir is in
the vacuum and $\bar N=0$ {\cite{paper2}}. In this case, the Lagrange
equation solutions ${\cal F}$ of (\ref{ndij}) for the three capacities
are functions only of $\omega_j/\Omega$ and $\eta$. Thus,
Eq.~(\ref{cidiefin}) gives
  \begin{eqnarray}
C_E&=&{\cal T}\;R_C\;{\cal C}(\eta)\;,\labell{celoss}\\
C&\geqslant&{\cal T}\;R_C\;{\cal K}(\eta)\;,\labell{ccqloss}\\
Q&\geqslant&{\cal T}\;R_C\;{\cal Q}(\eta)\;,\labell{qloss}
\end{eqnarray}
where the functions $\cal C$, $\cal K$ and $\cal Q$ take the place of
$\cal W$ in Eq.~(\ref{calci}) by replacing $w$ with $c_E$, $k$ and
$q$. These functions are plotted in Fig.~\ref{f:loss}.  In all these
three cases, the dependence on the input power is given by the
$\sqrt{\cal P}$ term in $R_C$. The alternative lower bound for $Q$ of
Eq.~(\ref{maggioraz}) becomes
  \begin{eqnarray} Q\geqslant{\cal T}\;R_C\;[{\cal C}(\eta)-1]
\;\labell{se}.
\end{eqnarray}
Both this bound and the function ${\cal Q}(\eta)$ of Eq.~(\ref{qloss})
are positive only for $\eta>1/2$. This reflects the fact that for
$\eta\leqslant 1/2$ the quantum capacity $Q$ is null. A simple
argument based on the no-cloning theorem {\cite{noclon}} is sufficient
to prove this, as in the case of the erasure channel {\cite{erasure}}.
In fact, assume that $Q$ is positive for $\eta<1/2$, and suppose that
a third party collects all the photons lost during the transmission:
to him the channel would appear to have a quantum efficiency
($1-\eta>1/2$) greater than the one of the receiver. If $Q>0$, both he
and the receiver would be able to reconstruct the quantum information
sent through the channel reliably, thus violating the no-cloning
theorem. Interestingly this $\eta=1/2$ bound for the lossy channel has
been observed also in tomographic quantum state reconstruction, where
the effect of the loss can be deconvolved from the reconstruction only
when the detection efficiency is bigger than $1/2$ {\cite{tomog}}.

Notice that, while $\cal C$ and $\cal Q$ must be computed through
numerical methods (a partial analytic characterization of   ${\cal
C}(\eta)$ is provided in {\cite{paper2}}, where, in particular, it is
shown that ${\cal C}(1)=2{\cal C}(1/2)=2$), it is possible to
determine analytically the value of $\cal K$. In fact, the Lagrange
equation for $C$ (see table~\ref{t:lagr}) has solution
\begin{eqnarray} 
N_j={\cal F}\left(\frac{\omega_j}\Omega,\eta_j\right)=
\frac{1/\eta_j}{e^{\omega_j/(\Omega{\eta_j})}-1}
\;\labell{ssss}.
\end{eqnarray}
When all the $\eta_j$ are equal, the value of $\Omega$ can be
calculated directly through the energy constraint (\ref{omegagr})
($y_0=0$ since there is no characteristic frequency $\bar\omega$). In
particular, since the function $f$ of Eq.~(\ref{defdif}) is
  \begin{eqnarray} f(\eta)=\int_0^\infty
dx\;\frac{x/\eta}{e^{x/\eta}-1}=\eta\;\frac{\pi^2}6 \;\labell{effe},
\end{eqnarray}
we have $\Omega=[12{\cal P}/({\pi\hbar\eta})]^{1/2}$. To evaluate
$\cal K$ through Eq.~(\ref{calci}), we need also the term
    \begin{eqnarray} \int_0^\infty dx\;k({\cal
F}(x,\eta),\eta)&=&\eta\int_0^\infty dy \;g\left(\frac
1{e^y-1}\right)\nonumber\\&=&\frac{\pi^2\eta}{3\ln 2} \;\labell{la}.
\end{eqnarray}
Replacing (\ref{effe}) and (\ref{la}) in Eq.~(\ref{calci}), we finally
find
\begin{eqnarray} {\cal K}(\eta)=\sqrt\eta\;\labell{re}.
\end{eqnarray}
Notice that, from Eq.~(\ref{classicalc}) it follows that in
Eq.~(\ref{ccqloss}) the equality must hold for $\eta=1$ and the
results of {\cite{caves,yuen}} are reobtained.

\begin{figure}[hbt]
\begin{center}
\epsfxsize=.7
\hsize\leavevmode\epsffile{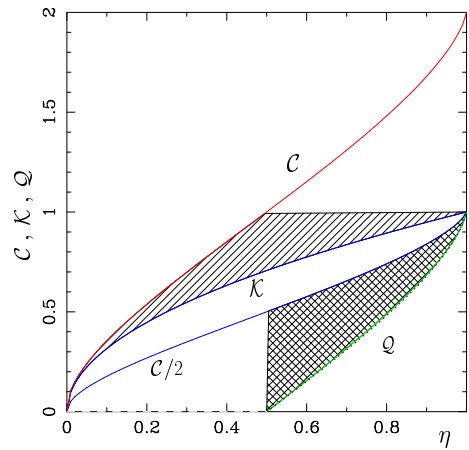}
\end{center}
\vspace{-.5cm}
\caption{Plot of the functions ${\cal C}(\eta)$, ${\cal K}(\eta)$ and
  ${\cal Q}(\eta)$ of Eqs.~(\ref{celoss})--(\ref{qloss}) for the lossy
  channel. The function ${\cal C}(\eta)$ characterizes the
  entanglement assisted capacity $C_E$ (upper continuous line).
  ${\cal C}(\eta)$, ${\cal K}(\eta)$ and the value 1 (that corresponds
  to the noiseless classical capacity (\ref{classicalc})) bound $C$
  which is restricted to the hatched region.  ${\cal C}(\eta)/2$ and
  ${\cal Q}(\eta)$ bound the quantum capacity $Q$ which is restricted
  to the cross-hatched region. The alternative lower bound for $Q$ of
  Eq.~(\ref{se}) is given by the dotted line, which is almost
  indistinguishable from ${\cal Q}(\eta)$. The quantum capacity $Q$ is
  null for $\eta<1/2$ (dashed line). The fact that, for the lossy
  channel ${\cal C}(1)=2$ is a signature of the superdense coding
  effect: prior entanglement allows to double the channel capacity
  {\cite{sdc}}.  } \labell{f:loss}\end{figure}

\section{White noise channel}\labell{s:white}
As in the case of the lossy channel, also the white noise channel has
no characteristic frequency $\bar\omega$, in fact all noise modes
contain the same average number of photons $\bar N_j=\bar N$. This
means that the solutions $\cal F$ of the Lagrange equation are only
functions of $\omega_j/\Omega$, $\bar N$ and $\eta$. Thus
Eq.~(\ref{cidiefin}) gives
\begin{eqnarray}
C_E&=&{\cal T}\;R_C\;{\cal C}(\bar N,\eta)\;,\labell{cew}\\
C&\geqslant&{\cal T}\;R_C\;{\cal K}(\bar N,\eta)\;,\labell{ccqw}\\
Q&\geqslant&{\cal T}\;R_C\;{\cal Q}(\bar N,\eta)\;,\labell{qw}
\end{eqnarray}
while Eq.~(\ref{maggioraz}) becomes   $Q\geqslant{\cal T}\;R_C\;[{\cal
C}(\bar N,\eta)-1]$. As in the previous case,
Eqs.~(\ref{cew})--(\ref{qw}) display the square root dependence of the
capacities on the input power $\cal P$ through $R_C$. The functions
$\cal C$, $\cal K$ and $\cal Q$ for the white noise channel are
plotted in Fig.~\ref{f:w} for different values of $\bar N$. Some
examples of the numerical solutions ${\cal F}(\omega_i/\Omega,\bar
N,\eta)$ of the Lagrange equations are plotted in Fig.~\ref{f:wsp}.

\begin{figure}[hbt]
\begin{center}
\epsfxsize=1.
\hsize\leavevmode\epsffile{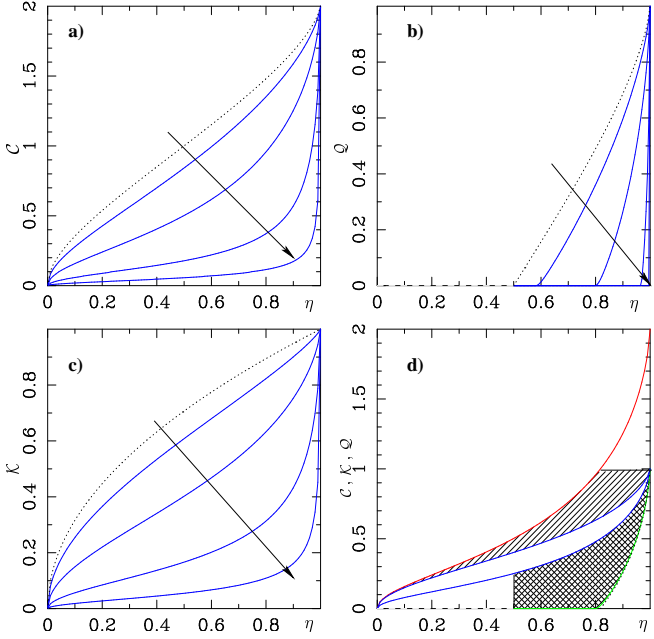}
\end{center}
\vspace{-.5cm}
\caption{{\bf a)}, {\bf b)}, {\bf c)} Plot of the functions ${\cal
    C}(\bar N,\eta)$, ${\cal K}(\bar N,\eta)$ and ${\cal Q}(\bar
  N,\eta)$ of Eqs.~(\ref{cew})--(\ref{qw}) for the white noise channel
  with $\bar N$ increasing along the direction of the arrows.  The
  dotted lines represents the case $\bar N=0$ from Fig.~\ref{f:loss}.
  {\bf d)}~Plot of the bounds of the classical (hatched region) and
  quantum (cross-hatched region) capacities for $\bar N=1$. From top
  to bottom, the curves are $\cal C$, $\cal K$, ${\cal C}/2$, $\cal Q$
  and the alternative lower bound ${\cal C}-1$, which in this case is
  practically coincident with $\cal Q$.  } \labell{f:w}\end{figure}

\begin{figure}[hbt]
\begin{center}
\epsfxsize=.6
\hsize\leavevmode\epsffile{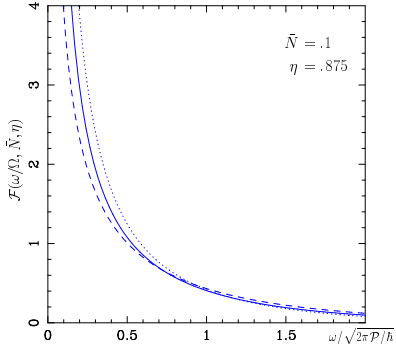}
\end{center}
\vspace{-.5cm}
\caption{Examples of the solution $N_j={\cal F}(\omega/\Omega,\bar
  N,\eta)$ of the Lagrange equations for $C_E$ (continuous line), for
  $C$ (dotted line from Eq.~(\ref{sol3})), and for $Q$ (dashed line).
} \labell{f:wsp}\end{figure}

As for the lossy channel, also here an analytical expression for $\cal
K$ exists. In fact, the Lagrange equation for $C$ (see
table~\ref{t:lagr}) has solution
\begin{eqnarray}
N_j={\cal F}\left(\frac{\omega_j}\Omega,\bar N,\eta_j\right)=
\frac {1/\eta_j}{e^{\omega_j/(\eta_j\Omega)}-1}-
\frac {1-\eta_j}{\eta_j}\bar N
\;\labell{sol3}.
\end{eqnarray}
Since $N_j$ represents the average photon number in the $j$th mode,
the solution (\ref{sol3}) can be used only when $N_j\geqslant 0$.
This condition is satisfied only if the frequency of the mode
$\omega_j$ is lower than the cut-off frequency
$\omega_{max}=\eta\;\Omega\;s$, where   \begin{eqnarray}
s\equiv\ln\left[1+\frac 1{(1-\eta)\bar N}\right]\;\labell{defs}
\end{eqnarray}
 (again we have assumed $\eta_j=\eta$ for all
$j$). For $\omega_j\geqslant\omega_{max}$, Eq.~(\ref{sol3}) cannot be
used (it gives a negative $N_j$) so that we assume $N_j=0$. This
physically corresponds to not sending any photons in the high
frequency modes: it would be too expensive in energetic terms to
contrast the noise in these modes.  With this choice, from
Eq.~(\ref{defdif}) we obtain \begin{eqnarray}
f(\bar N,\eta)&=&\int_0^{\omega_{max}/\Omega}
dx\;x\;\left[\frac{1/\eta}{e^{x/\eta}-1}-\frac{1-\eta}\eta\bar
  N\right]\nonumber\\
&=&\eta\left[\Gamma(s)-(1-\eta)s^2\bar N/2\right]
\;\labell{defdif2},
\end{eqnarray}
where
 \begin{eqnarray}
\Gamma(x)\equiv\int_0^xdy\;\frac y{e^y-1}
\;\labell{defgamma}.
\end{eqnarray}
To calculate $\cal K$ we also need the integral \begin{eqnarray}
&&\int_0^\infty dx\;k\left({\cal F}(x,\bar N,\eta),\bar
  N,\eta\right)\nonumber\\&=&
\int_0^{\eta s}dx\;\left[g\left(\eta{\cal F}+(1-\eta)\bar
    N\right)-g((1-\eta)\bar N)\right]\nonumber\\&=&
\eta\left[\int_0^{s}\!\!  dx\;g\left(\frac
      1{e^x-1}\right)-s\;g((1-\eta)\bar N)\right].
\;\labell{kappac}
\end{eqnarray}
Replacing (\ref{defdif2}) and (\ref{kappac}) into Eq.~(\ref{calci})
the function ${\cal K}(\bar N,\eta)$ is determined.  Notice that in
the limit $\bar N\to 0$, it is possible to show that the function
$\cal K$ converges to $\sqrt\eta$ so that one reobtains the results of
the lossy channel.

\section{Thermal noise}\labell{s:thce}
Let us now analyze the case of thermal noise. This noise model does
have a characteristic frequency $\bar\omega\equiv KT/\hbar$ which
depends on the bath temperature $T$. Since $\bar\omega\neq 0$, we need
to solve $y_0$ from Eq.~(\ref{y}) which clearly implies that $y_0$ is
a function of the ratio between the square of the temperature $T$ and
the power $\cal P$. In this case, $\bar N=1$ and the expressions for
the capacities are \begin{eqnarray}
  C_E&=&{\cal T}\;R_C\;{\cal C}(y_0^{C_E},\eta)\;,\labell{ceth}\\
  C&\geqslant&{\cal T}\;R_C\;{\cal K}(y_0^C,\eta)\;,\labell{ccqth}\\
  Q&\geqslant&{\cal T}\;R_C\;{\cal Q}(y_0^Q,\eta)\;,\labell{qth}
\end{eqnarray}
where $y_0^{C_E}$,$y_0^{C}$ and $y_0^{Q}$ are the solutions of
Eq.~(\ref{y}) for the respective capacities.  Moreover, the
alternative lower bound of Eq.~(\ref{maggioraz}) gives
$Q\geqslant{\cal T}\;R_C\;[{\cal C}(y_0^{C_E},\eta)-1]$. The presence
of the terms $y_0$ complicates the dependence of the capacities on the
input power $\cal P$. However, once the ratio $\hbar(KT)^2/{\cal P}$
has been fixed, the usual dependence on the square root of the input
power applies. Some numerical plots of ${\cal C}$, $\cal K$ and $\cal
Q$ are given in Fig.~\ref{f:ther} as a function of $\eta$ and of the
temperature $T$. Some examples of the corresponding Lagrange equation
solutions ${\cal F}$ are given in Fig.~\ref{f:spth}.

\begin{figure}[hbt]
\begin{center}
\epsfxsize=1.
\hsize\leavevmode\epsffile{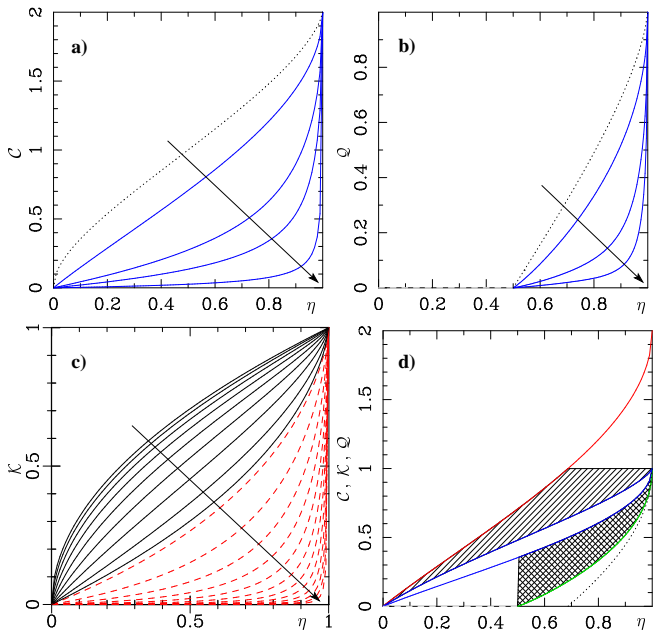}
\end{center}
\vspace{-.5cm}
\caption{{\bf a)}, {\bf b)}, {\bf c)}~Plot of the functions ${\cal
    C}(y_0^{C_E},\eta)$, ${\cal K}(y_0^{C},\eta)$ and ${\cal
    Q}(y_0^Q,\eta)$ of Eqs.~(\ref{ceth})--(\ref{qth}) for the thermal
  channel with temperature $T$ increasing along the direction of the
  arrows. The dotted lines represents the case $T=0$ from
  Fig.~\ref{f:loss}. In the plot {\bf c)}, the low temperature regime
  $T<T_c$ (continuous lines) is obtained from Eq.~(\ref{calkth}),
  while the high temperature regime (dashed lines) is obtained from
  Eq.~(\ref{bfin2}). {\bf d)}~Plot of the bounds of the classical
  (hatched region) and quantum (cross-hatched region) capacities for
  $R_T/R_C=.41$. From top to bottom, the curves are $\cal C$, $\cal
  K$, ${\cal C}/2$, $\cal Q$ and the alternative lower bound ${\cal
    C}-1$.  } \labell{f:ther}\end{figure}

\begin{figure}[hbt]
\begin{center}
\epsfxsize=.6
\hsize\leavevmode\epsffile{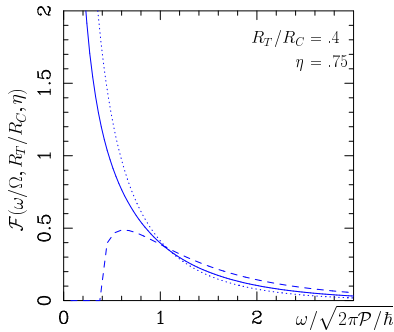}
\end{center}
\vspace{-.5cm}
\caption{Examples of the solution $N_j={\cal F}(\omega/\Omega,\bar
  N,\eta)$ of the Lagrange equations for $C_E$ (continuous line), for
  $C$ (dotted line from Eq.~(\ref{sol3})), and for $Q$ (dashed line).
  It is possible to show that the solutions of the Lagrange equation
  for $Q$ have two cut-off frequencies for low and high $\omega$. In
  this graph only the first one is evident.  }
\labell{f:spth}\end{figure}

Again the function $\cal K$ for the lower bound of the classical
capacity can be solved analytically. We will find that below a
critical temperature $T_c$ the solutions $N_j$ of the Lagrange
equation, i.e.
\begin{eqnarray} N_j={\cal
    F}\left(\frac{\omega_j}\Omega,\frac{\bar\omega}\Omega,\eta_j\right)=
  \frac {1/\eta_j}{e^{\omega_j/(\eta_j\Omega)}-1}- \frac
  {(1-\eta_j)/\eta_j}{e^{\omega_j/\bar\omega}-1} \;,\labell{sol4}
\end{eqnarray}
are valid for all frequencies $\omega_j$. On the contrary, when
$T>T_c$ a cut-off frequency arises above which (as in the case of the
white noise channel) it is convenient not to send photons.

In the low temperature regime we find \begin{eqnarray}
y^C_0=\frac{\eta\;R_T}{[\eta\;R_C^2+(1-\eta)R_T^2]^{1/2}}\;\labell{ylt},
\end{eqnarray} 
where $R_T\equiv \frac{\pi^2}{3\ln 2}\frac{KT}{h}$, so that the
Lagrange multiplier $\Omega$ obtained from the energy constraint
(\ref{omegagr}) is
    \begin{eqnarray} \Omega=\frac {6\ln 2}{\eta\;\pi}\sqrt{\eta
R_C^2+(1-\eta)R_T^2} \;\labell{om}.
\end{eqnarray}
Replacing Eqs.~(\ref{sol4}) and (\ref{om}) in Eq.~(\ref{calci}), we
obtain \begin{eqnarray}
{\cal K}(y_0^C,\eta)=\sqrt{\eta+(1-\eta)\left(\frac{R_T}{R_C}\right)^2}-
\frac{\Lambda(1-\eta)}{\Lambda(1)}\frac{R_T}{R_C}
\;\labell{calkth},
\end{eqnarray}
where the dependence on $y_0^C$ derives from (\ref{ylt}) and where
\begin{eqnarray}
\Lambda(x)\equiv\ln 2\int_0^\infty dy\;g\left(\frac x{e^y-1}\right)
\;\labell{lambdagr}.
\end{eqnarray}
Equations (\ref{sol4}) and (\ref{om}) are consistent (i.e. provide a
nonnegative $N_j$ for all frequencies) only in the low temperature
regime $T\leqslant T_c\equiv\sqrt{6{\cal P}}/(\pi K)$, where
$R_C\geqslant R_T$ and the $N_j$'s of Eq.~(\ref{sol4}) are positive
quantities for all $j$. On the other hand, in the high temperature
regime ($T>T_c$), the solutions $N_j$ provided by Eq.~(\ref{sol4}) are
valid only for frequencies
$\omega_j\leqslant\omega_{max}\equiv\eta\;KT(\ln\xi)/(\hbar\;y_0^C)$,
where the parameters $y_0^C$ and $\xi$ are obtained by solving (for
$\xi>1$) the following coupled equations (the first is determined by
imposing $N_j=0$ in Eq.~(\ref{sol4}), while the second is just
Eq.~(\ref{y})):
\begin{eqnarray}
&&\xi^{\eta/y_0^C}-(1-\eta)\xi-\eta=0
  \;\labell{eqq}
\\\nonumber&&
      \eta\;\Gamma(\ln\xi)=
\left[\frac{1-\eta}\eta\;\Gamma\!\left(\frac{\eta\ln\xi}{y_0^C}\right)+
\frac{\pi^2}6
      \left(\frac{R_C}{R_T}\right)^2\right](y_0^C)^2\;,
\end{eqnarray}
with $\Gamma(x)$ defined in Eq.~(\ref{defgamma}).  For higher
frequencies Eq.~(\ref{sol4}) gives negative values of $N_j$ and we
need to choose $N_j=0$. With these solutions $N_j$, one can evaluate
$y_0^C$ and $\Omega$ through Eqs.~(\ref{y}) and (\ref{omegagr}) so
that from Eq.~(\ref{calci}) it is possible to obtain
\begin{eqnarray}
&&{\cal K}(y_0^C,\eta)\labell{bfin2}=
\frac{3\;\eta\;\ln 2}{\pi^2\;y_0^C}\frac{R_T}{R_C}
\\&&\nonumber\times
\int_0^{\ln\xi}{dx}
\left[g\left(\frac
1{e^{x}-1}\right)-g\left(\frac{1-\eta}{e^{x\;\eta/y_0^C}-1}\right)\right]
\;.
\end{eqnarray}
In the limit $T\to T_c$, we find $y_0^C\to\eta$ and $\xi\to\infty$, and
Eq.~(\ref{bfin2}) reduces to (\ref{calkth}), i.e. the transition
between the high temperature regime and the low temperature regime is
continuous.  On the other hand, for $T=0$, we find $R_T=0$, so from
Eq.~(\ref{calkth}) we reobtain the lossy channel result of
Eq.~(\ref{re}). The bounds (\ref{calkth}) and (\ref{bfin2}) are
plotted in Fig.~\ref{f:ther}c for different values of $T$.

\begin{table}\begin{tabular}{|c|c|}
\hline&\\$C_E$&$\displaystyle\left(1+\frac
1{N_j}\right)^2=e^{\omega_j/\Omega}\left (1+\frac
2{\tilde D_j-1}\right)^{\frac{2(1-\eta_j)(2N_j+1)}{\tilde D_j}}$ \\ & \\ \hline 
$C$&\\ {\scriptsize lower } & $\displaystyle\left(1+\frac
1{N_j}\right)=e^{\omega_j/\Omega}\left(1+\frac
1{(1-\eta_j)N_j}\right)^{1-\eta_j}$
\\{\scriptsize bound}& \\ \hline$Q$ &\\
{\scriptsize lower} & $\displaystyle\left(1+\frac
    1{N_j}\right)=e^{\omega_j/\Omega}\left (1+\frac
    2{\tilde D_j-1}\right)^{\frac{2(1-\eta_j)(2N_j+1)}{\tilde D_j}}$ \\{\scriptsize 
bound}&\\ \hline  \end{tabular}\caption{Lagrange equations for the
dephasing channel deriving
from (\ref{lagr}) for the different capacities. 
\labell{t:lagrdep}} 
\end{table}

\section{Dephasing channel}\labell{s:depch}
In this section we will focus on a non-linear noise source where the
effect of the reservoir depends on the state of the message. In
particular, we consider the case in which the average photon number of
the noise source $\bar N_j$ is the same as the one ($N_j$) of the
message. The average photon number is hence preserved during the
transmission. Of course, this does not mean that the channel is immune
to noise: in replacing the lost photons with those from the reservoir,
some phase correlations are lost. Under these conditions, imposing
$\bar N_j=N_j$ in (\ref{cdiewhite}), (\ref{dk}) and (\ref{dq}), the
values of the functions $c_E$, $k$ and $q$ become
\begin{eqnarray} 
c_E(N_j,\eta_j)&=&2[g(N_j)-g((\tilde D_j-1)/2)]\;\labell{ced}\\
k(N_j,\eta_j)&=&g(N_j)-g((1-\eta_j)N_j)\;\labell{ccqd}\\
q(N_j,\eta_j)&=&g(N_j)-2g((\tilde D_j-1)/2)\;\labell{qd},
\end{eqnarray}
with $\tilde D_j\equiv\sqrt{1+4N_j(N_j+1)(1-\eta_j)}$, from which the
Lagrange equations of table~\ref{t:lagrdep} derive to replace those of
table~\ref{t:lagr}.  As in the case of the lossy channel the solutions
of the Lagrange equations depend only on $\omega_j/\Omega$ and $\eta$.
Hence, the same structure as Eqs.~(\ref{celoss})--(\ref{qloss})
applies, but here the functions $\cal C$, $\cal K$ and $\cal Q$ are
the ones plotted in Fig.~\ref{f:dep}.

\begin{figure}[hbt]
\begin{center}
\epsfxsize=.7
\hsize\leavevmode\epsffile{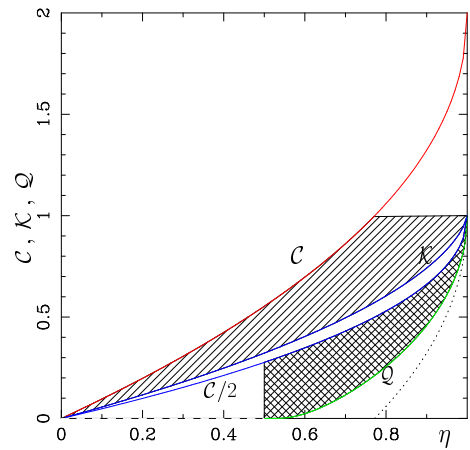}
\end{center}
\vspace{-.5cm}
\caption{Plot of the functions ${\cal C}(\eta)$, ${\cal K}(\eta)$ and
  ${\cal Q}(\eta)$ for the dephasing channel. The hatched region and
  the cross-hatched region are defined as in Fig.~\ref{f:loss} and
  contain the classical and quantum capacities respectively. The
  alternative lower bound for $Q$ (i.e. ${\cal C}(\eta)-1$) is given
  by the dotted line.  As for the other channels, it can be shown that
  $Q$ is null for $\eta<1/2$ (dashed line). }
\labell{f:dep}\end{figure}

\section{Conclusions}\labell{s:conc}
In this paper we extend previous analysis on the capacities of
broadband bosonic channels with input power constraint
{\cite{caves,yuen}}.  In particular, we analyzed the quantum
capacities in the presence of different noise sources.  Solutions for
the entanglement assisted capacity $C_E$ and upper and lower bounds
for the classical and quantum capacities $C$ and $Q$ were provided. At
least in the case of unit quantum efficiency (i.e. when the channel is
noise free), these bounds are tight since they reproduce the noiseless
capacities {\cite{caves,yuen}}. Moreover, if the channel noise does
not have any characteristic frequency (as in the case of the loss,
white noise and dephasing), the square root of the input power
dependence (that was known for the noiseless case) is reobtained.
Even though all the results in the paper were obtained by considering
a uniform quantum efficiency for all the channel modes, the procedure
can be extended also to non uniform configurations.  It is also
possible to include frequency degerate situations, e.g. where one uses
polarization degrees of freedom to encode information.  It is still to
be determined whether non-linearities in the channel dynamics (where
some known interaction couples different modes) can be used
{\cite{caves}} to beat the square root dependence, as in the case of
the qubit channel discussed in {\cite{seth}}.

\acknowledgements V. G. and L. M.  thank Prof. G. M. D'Ariano for
useful discussion and comments.  This work was funded by the ARDA,
NRO, NSF, and by ARO under a MURI program.

\appendix
\section{Single mode entropies}\labell{s:entr}
In this appendix we calculate some relevant entropic quantities for
the single modes when the input is a Gaussian state. We follow the
derivation of Holevo and Werner {\cite{holevo}} and, for ease of
notation, the mode index $j$ is dropped.

\paragraph*{Quantum mutual information.---} The quantum mutual
information $I({\cal N},\varrho)$ (\ref{qmi}) for a single mode can be
evaluated just considering the correlation matrix $\alpha$ of the mode
input state $\varrho$, defined as
\begin{eqnarray}
\alpha=\left[\matrix{\langle\Delta
q^2\rangle&\frac 12\langle\{\Delta q,\Delta
p\}\rangle\cr\cr\frac 12\langle\{\Delta p,\Delta
q\}\rangle&\langle\Delta p^2\rangle}\right]
\;\labell{alpha},
\end{eqnarray}
where $\{\cdot,\cdot\}$ denotes the anticommutator, $\Delta q\equiv
q-\langle q\rangle$ and $\Delta p\equiv p-\langle p\rangle$, with $q$
and $p$ the two orthogonal quadratures
$q\equiv\sqrt{\hbar/2}\;(a^\dag+a)$ and $p\equiv
i\sqrt{\hbar/2}\;(a^\dag-a)$. In {\cite{holevo}} it has been shown
that, for a given value of the matrix $\alpha$, $I({\cal N},\varrho)$
achieves its maximum value for the Gaussian state
\begin{eqnarray}
\varrho=\frac\hbar{2\pi}\int dz\;\exp\left[-i\;z\cdot
\left(\matrix{\Delta q\cr\Delta
p}\right)- \frac{z\cdot\alpha\cdot z^T}2\right]
\;\labell{gaus},
\end{eqnarray}
where $z$   is a real bidimensional line vector.  According to
Eq.~(\ref{qmi}), to determine the value of $I({\cal N},\varrho)$ we
need the evaluate the input, output and exchange entropies. Following
{\cite{holevo,hirota,holevo2} the input entropy of the Gaussian state
$\varrho$ can be calculated as   \begin{eqnarray} S(\varrho)=
g(\sqrt{\lambda_+\lambda_-}-1/2) \;\labell{ent1},
\end{eqnarray}
where the function $g$ is defined in (\ref{defdig}) and $\lambda_\pm$
are the eigenvalues of $\alpha/\hbar$.  In the same way, we can
evaluate also the final entropy $S({\cal N}[\varrho])$. In fact, the
state ${\cal N}[\varrho]$ (evolved by the map $\cal N$ defined in
(\ref{ch})) is again Gaussian and has correlation matrix
\begin{eqnarray}
\alpha'=\eta\:\alpha+(1-\eta)\;B
\;\labell{alphaprimo},
\end{eqnarray}
where $B$ is the correlation matrix of the $j$th noise mode introduced
in (\ref{matriceb}). This means that \begin{eqnarray} S({\cal
    N}[\varrho])= g(\sqrt{\lambda'_+\lambda'_-}-1/2)
  \;\labell{ent2},\end{eqnarray} where $\lambda'_\pm$ are the
eigenvalues of $\alpha'/\hbar$. The calculation of the entropy of
exchange requires to specify a purification $\Phi_\varrho$ of
$\varrho$: a good choice is the two-mode Gaussian state
  \begin{eqnarray} &&\Phi_{\varrho}= \left(\frac\hbar{2\pi}\right)^2\int
dz\int d\bar z\;\exp \Big[-i(\Delta q,\Delta
p)\cdot z^T\nonumber\\
&&-i(\Delta\bar q,\Delta\bar p)\cdot\bar z^T- (z,\bar z)\cdot M\cdot
(z,\bar z)^T/2\Big] \;\labell{purif},
\end{eqnarray}
where the $\bar q$ and $\bar p$ are quadratures operators acting on an
ancillary mode and the 4$\times$4 two-mode correlation matrix $M$
is\begin{eqnarray} M&=&\left[\matrix{\alpha&\beta
      \cr-\beta&\alpha}\right] \;,\;\labell{alphagrande}
\end{eqnarray}
with $\beta=\Delta\sqrt{-(\Delta^{-1}\alpha)^2-\frac\openone 4}$,
$\Delta$ being the $2\times 2$ matrix $\hbar\left[ \matrix{0&1\cr-1&0}\right]$.
The map ${\cal N}\otimes\openone$ evolves $\Phi_\varrho$ into a
Gaussian state of the same form of (\ref{purif}) with correlation
matrix
\begin{eqnarray}
&&M'=
\left[\matrix{\alpha'&\sqrt{\eta}\beta\cr-
\sqrt{\eta}\beta&\alpha}\right]
\;.\;\labell{aprimo}
\end{eqnarray} 
According to {\cite{holevo,hirota,holevo2}} the entropy of exchange
can be calculated as     \begin{eqnarray} S(({\cal
N}\otimes\openone)[\Phi_\varrho]) =\frac
12\sum_{k=1}^4g(|\lambda_k|-1/2)\;,\labell{entex}
\end{eqnarray}where $\lambda_{1},\cdots,\lambda_4$ are eigenvalues of the matrix
$\Delta_{12}^{-1}M'/\hbar$, $\Delta_{12}$ being the $4\times 4$ matrix
$\left[\matrix{\Delta&0\cr0&-\Delta}\right]$.

\begin{figure}[t]
\vspace{.5cm}
\begin{center}
\epsfxsize=.7
\hsize\leavevmode\epsffile{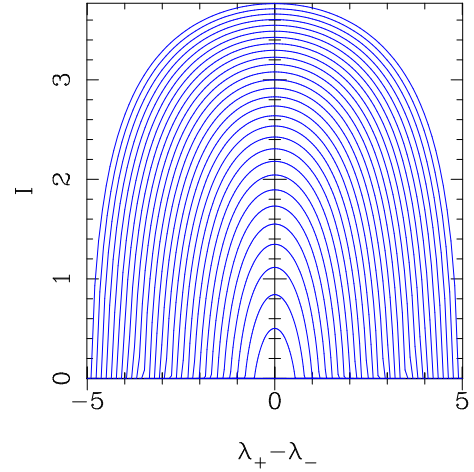}
\end{center}
\vspace{-.5cm}
\caption{Plot of the quantum mutual information $I({\cal N},\varrho)$
  of the single mode as a function of $\lambda_+-\lambda_-$. The
  different curves correspond to different values of the eigenvalues
  sum $\lambda_++\lambda_-$ increasing from bottom to top. It is
  evident that the maxima are always achieved for
  $\lambda_+=\lambda_-$. The parameters for these plots are $\bar
  N=.1$; $\eta=.8\;$.}  \labell{f:entropy}\end{figure}

In order to evaluate the expressions for the entropies, it is
convenient to introduce the following real parametrization:
      \begin{eqnarray} \alpha=\frac\hbar 2 \left[\matrix{n_0e^r&c\cr
c&n_0e^{-r}}\right] \;\labell{nuovoalpha},
\end{eqnarray}
where $r$ is the squeezing parameter. These parameters are related
through the average number of photons $N$ by the inequalities
$\sqrt{c^2+1}\leqslant n_0=(2N+1-m)/\cosh r$ (with $m=\langle
q/\hbar\rangle^2+\langle p/\hbar\rangle^2$): the first relation
derives from the Heisenberg uncertainty relation, while the second
from the energy constraint. The eigenvalues of $\alpha$ and $\alpha'$
are respectively
  \begin{eqnarray} \lambda_\pm&=&\frac  
12[n_0\cosh(r)\pm\sqrt{(n_0\sinh(r))^2+c^2}]\;\labell{avala},\\
\lambda_\pm'&=&\eta\lambda_\pm+(1-\eta)
\hbar(\bar N+1/2)\;\labell{avalap},
\end{eqnarray}
while the four eigenvalues of the matrix $\Delta_{12}^{-1}M'/\hbar$
are $\lambda_{1,2,3,4}=\pm[(L_0\pm\sqrt{L_1+L_0^2})/8]^{1/2}$, where
\begin{widetext}\begin{eqnarray} L_0&=&-(1+\eta^2)-4(1-\eta)^2\bar
    N^2-4(1-\eta)[1-\eta+\eta(\lambda_++\lambda_-)]\bar N
    -2\eta(1-\eta)(\lambda_++\lambda_-)-4(1-\eta)^2\lambda_+\lambda_-
    \;\labell{elle0}\\L_1&=&-8(1-\eta)(1+2\bar N)[2(1-\eta)(1+2\bar
    N)\lambda_+\lambda_-+\eta(\lambda_++\lambda_-)]-4\eta^2  \labell{elle1}.
\end{eqnarray}\end{widetext}
From these equations and from the definition (\ref{qmi}) it is evident
that the entropies of Eqs.~(\ref{ent1}) depend on the parameters
$n_0$, $r$ and $c$ only through the eigenvalues $\lambda_\pm$ of
$\alpha$. Since these quantities are related with the average number
of photons $N$ by   \begin{eqnarray} \lambda_++\lambda_-=2N+1-m
\;\labell{en},
\end{eqnarray}
one can show that the maximum of $I({\cal N},\varrho)$ for fixed $N$
and $m$ is obtained for $\lambda_+=\lambda_-$ (see
Fig.~\ref{f:entropy}). According to Eq.~(\ref{avala}), this is
equivalent to requiring $r=0$ (no squeezing) and $c=0$ (maximally
mixed states). Imposing $\lambda_+=\lambda_-$ and maximizing with
respect to $m$ in the above relations, one easily finds that the
entropies become
\begin{eqnarray} S(\varrho)&=&g(N)\;\labell{entropie1}\\ S({\cal
N}[\varrho])&=&g(N')\;\labell{entropie2}\\ S(({\cal
N}\otimes\openone)[\Phi_\varrho])&=&g\left(\frac{D+N-N'-1}2\right)
\labell{entropie3}
\\\nonumber&&+
g\left(\frac{D-N+N'-1}2\right)
\;,
\end{eqnarray}
where $N'$ and $D$ are defined in (\ref{nprimo}) and (\ref{dj}). From
these relations it is immediate to show that the maximum of the
quantum mutual information for a given value of the average photon
number $N$ is given by Eq.~(\ref{cdiewhite}).

\paragraph*{Coherent information.---} Replacing 
(\ref{entropie1})--(\ref{entropie3}) in Eq.~(\ref{coher}) allow us to
calculate the value of the coherent information $J({\cal N}, \varrho)$
for Gaussian states of Eq.~(\ref{gaus}) with no squeezing and $m=0$.
This gives the function $q$ of Eq.~(\ref{dq}).

\paragraph*{Holevo quantity.--} To calculate the Holevo quantity $\chi_{\cal N}$ for the code
introduced in Sec.~\ref{s:ccq} we can use the above results. In fact
both the global state $\varrho$ and its components $\sigma(\mu)$ are
Gaussian states of the form (\ref{gaus}), with correlation matrices
$\hbar (N+1/2)\openone$ and $\hbar\;\openone/2$ respectively. Hence it
is immediate to calculate the entropies that allow to obtain the value
of $\chi_{\cal N}$ reported in Eq.~(\ref{dk}).

\end{document}